\documentclass[12pt]{article}
\usepackage{cite,psfig}

\def\be{\begin{equation}}
\def\ee{\end{equation}}
\def\bea{\begin{eqnarray}}
\def\eea{\end{eqnarray}}

\renewcommand{\phi}{\varphi}

\begin{document}

\thispagestyle{empty}
\vspace*{.5cm}
\noindent
{\large HD-THEP-00-22\hfill March 2000}\\
\vspace*{2.5cm}

\begin{center}
{\Large\bf Quintessential Adjustment\\[.3cm] 
of the Cosmological Constant}
\\[2.5cm]
{\large A. Hebecker and C. Wetterich}\\[.5cm]
{\it Institut f\"ur Theoretische Physik der Universit\"at Heidelberg\\
Philosophenweg 16, D-69120 Heidelberg, Germany}\\[2.1cm]

{\bf Abstract}\end{center}
\noindent
We construct a time dependent adjustment mechanism for the cosmological 
``constant'' which could be at work in a late Friedmann-Robertson-Walker 
universe dominated by quintessence and matter. It makes use of a 
Brans-Dicke field that couples to the evolving standard-model vacuum energy 
density. Our explicit model possesses a stable late-time solution with a 
fixed ratio of matter and field energy densities. No fine tuning of model 
parameters or initial conditions is required. 
\vspace*{2cm}
\newpage

\section{Introduction}
The extraordinary smallness of the observational upper bound on the 
cosmological constant conflicts with naive field theoretic expectations. 
This is a well-known fundamental problem in our understanding of the 
interplay between gravity and the known quantum field theories (see, 
e.g.,~\cite{wein} for a review). Recent observations~\cite{perl} suggesting 
a small non-zero value for the cosmological constant make this problem even 
more severe since an exact zero, possibly the result of a yet unknown 
symmetry, is replaced by a small number, $\epsilon_{vac}/M_p^4\simeq 
2\times 10^{-123}$. In any fundamental unified theory this number would 
have to be calculable. 

A possible alternative is a homogeneous contribution to the energy density 
of the universe which varies with time. It is typically connected to a time 
varying scalar field -- the cosmon -- which relaxes asymptotically for large 
time to zero potential $\mbox{energy~\cite{wet,rp}}$. The late time behavior 
in these models of ``quintessence''~\cite{cds} is insensitive to the initial 
conditions due to the stable attractor properties of the asymptotic 
solution. The homogeneous fraction of the cosmic energy density may be 
constant or slowly increase with time (say from 0.1 during nucleosynthesis 
to 0.7 today)~\cite{wet,as}, in sharp contrast to a cosmological constant, 
which needs to be adjusted such that it becomes important precisely today. 

It has been argued~\cite{wet} that the relaxation to a zero value of the 
effective potential rather than to a constant is connected to the dilatation 
anomaly. In absence of a fundamental theory it is, however, not obvious how 
to verify (or falsify) this assertion. We explore here an alternative 
possibility, namely that the value to which the effective cosmon potential 
relaxes is itself governed by a field which dynamically adjusts 
the ``cosmological constant'' to zero. We consider our proposal as an 
existence proof that such a mechanism can work. It is conceivable that 
simpler and more elegant models can be found once the basic adjustment 
mechanism is identified. 

Rubakov has recently suggested a mechanism for the dynamical adjustment 
of the cosmological constant to zero~\cite{rub} that avoids Weinberg's 
no-go-theorem~\cite{wein} in a very interesting way (cf.~\cite{dol} for 
other recent work on adjustment mechanisms). In his scenario, a scalar 
field governing the value of the cosmological constant rolls down a 
potential and approaches the zero of the potential, i.e., the point where 
the cosmological constant vanishes, as $t\to\infty$. Such behavior is 
realized by a diverging kinetic term~\cite{sa}, which depends on a second 
scalar field. This field, a Brans-Dicke-field that couples to the current 
value of the cosmological constant, ensures the stability of the solution. 

However, in Rubakov's model the universe is inflating after the adjustment 
of the cosmological constant. This makes it necessary to add a period of 
reheating, which implies the need to fine tune the minimum of the inflaton 
potential to zero. In this sense, the fine tuning problem for the total 
effective potential is now shifted to the inflaton sector. Furthermore, it 
is difficult to imagine testable phenomenological consequences of such an 
adjustment at inflation. 

The present paper suggests a dynamical adjustment mechanism for the 
cosmological constant that can be at work in a realistic, late 
Friedmann-Robertson-Walker universe. In this model, the energy density is 
dominated by non-standard-model dark matter together with a cosmon field 
$\phi$. This field can account for a homogeneous part of the total energy 
density which does not participate in structure formation and may lead to 
an accelerated expansion of the universe today. As in Rubakov's scenario, 
the cosmological constant, characterized by a field $\chi$, rolls down a 
potential and approaches zero asymptotically. This is realized by a kinetic 
term for $\chi$ that depends on $\phi$ and diverges as $t^4$ when $\phi(t) 
\to\infty$ at large $t$. To make this solution insensitive to changing 
initial conditions, a Brans-Dicke field $\sigma$ is introduced. This field 
`feels' the current value of the cosmological constant and provides the 
required feedback to the diverging kinetic term. 

Thus, a realistic, late cosmology with an asymptotically vanishing 
cosmological constant arises. Baryons can be added as a small perturbation 
and do not affect the stability of the solution. In the concrete numerical 
example provided below, their coupling to the Brans-Dicke field $\sigma$ is 
not yet realistic. 

In the following, the cosmological model outlined above is explicitly 
constructed.

\section{Adjusting a scalar potential to zero in a given 
Friedmann-Robertson-Walker background}
The action of the present model can be decomposed according to 
\be
S=S_E+S_{SF}+S_{SM}\,,
\ee
where $S_E$ is the Einstein action, $S_{SF}$ the scalar field action, and 
$S_{SM}$ the standard model action, which is written in the form
\be
S_{SM}=S_{SM}[\psi,g_{\mu\nu},\chi]=\int d^4x\sqrt{g}\,
{\cal L}_{SM}(\psi,g_{\mu\nu},\chi)\,.\label{ssm}
\ee
Here $g=-\mbox{det}(g_{\mu\nu})$ and $\psi$ stands for the gauge fields, 
fermions and non-singlet scalar fields of the standard model (or some 
supersymmetric or grand unified extension). The scalar singlet $\chi$ is 
assumed to govern the effective UV-cutoffs of the different modes of 
$\psi$, thereby influencing the effective cosmological constant. Units are 
chosen such that $M^2=(16\pi G_N)^{-1}=1$. 

Integrating out the fields $\psi$, one obtains (up to derivative terms)
\be
S_{SM}=\int d^4x\sqrt{g}\,V(\chi)\,.
\ee
Let the potential $V(\chi)$ have a zero, $V(\chi_0)=0$ with $\alpha=V'( 
\chi_0)$, and rename the field according to $\chi\,\to\,\chi_0+\chi$. Then 
the action near $\chi=0$ becomes 
\be
S_{SM}=\int d^4x\sqrt{g}\,\alpha\chi\,.
\ee
Due to this potential the field $\chi$ will decrease (for $\alpha>0$) during 
its cosmological evolution. It can be prevented from rolling through the 
zero by a diverging kinetic term~\cite{rub}. 

First, let the geometry be imposed on the system, i.e., assume a flat FRW 
universe with Hubble parameter $H=(2/3)\,t^{-1}$, independent of the 
dynamics of $\chi$. With a kinetic Lagrangian 
\be
{\cal L}_{SF}=\frac{1}{2}\,\partial^\mu\chi\partial_\mu\chi\,F(t)\,,
\ee
one obtains the following equation of motion for $\chi$:
\be
\ddot{\chi}+(3H+\dot{F}/F)\dot{\chi}+(\partial V/\partial \chi)/F=0\,. 
\label{eomc}
\ee
Here $F(t)=t^4$ is an externally imposed condition which will be realized 
below by the quintessence field, which serves as a clock. 
Equation~(\ref{eomc}) has the particular solution $\chi=(\alpha/6)\, 
t^{-2}$, which provides an acceptable late cosmology since all energy 
densities associated with $\chi$ scale as $t^{-2}$. 

Clearly, it requires fine tuning of the initial conditions to achieve 
the desired behavior $\chi\to 0$ as $t\to\infty$, which is realized in 
this particular solution. However, this fine tuning can be avoided by 
adding a Brans-Dicke field $\sigma$ that `feels' the deviation of $\chi$ 
from zero and provides the appropriate `feedback' to the kinetic term 
so that $\chi$ reaches zero asymptotically independent of its initial value. 

The field $\sigma$ has a canonical kinetic term and it is coupled to 
${\cal L}_{SM}$ by the substitution $g_{\mu\nu}\to g_{\mu\nu}\sqrt{\sigma}$ 
in Eq.~(\ref{ssm}), 
\be
S_{SM}=S_{SM}[\psi,g_{\mu\nu}\sqrt{\sigma},\chi]=\int d^4x\,\sigma\sqrt{g}\, 
{\cal L}_{SM}(\psi,g_{\mu\nu}\sqrt{\sigma},\chi)\,.\label{fco}
\ee
Integrating out the fields $\psi$, one obtains now 
\be
S_{SM}=\int d^4x\sqrt{g}\,\alpha\sigma\chi\label{ssmf}
\ee
near $\chi=0$. The scalar field Lagrangian is now taken to be 
\be
{\cal L}_{SF}=\frac{1}{2}(\partial\chi)^2F(\sigma,t)+\frac{1}{2}
(\partial\sigma)^2-\beta\sigma t^{-2}\,,\label{lsf}
\ee
where $F(\sigma,t)=\sigma^2t^4$. The additional $t$ dependence of the term 
$\sim\beta$ is again introduced ad hoc and will later on be realized by the 
dynamics of the cosmon. 

Now Eq.~(\ref{eomc}) is supplemented with the equation of motion for 
$\sigma$, 
\be
\ddot{\sigma}+3H\dot{\sigma}+\alpha\chi-\beta t^{-2}-\sigma\dot{\chi}^2t^4= 
0\,.
\ee
The combined equations have the asymptotic solution $\chi=\chi_0\,t^{-2}$ 
and $\sigma=\sigma_0=\mbox{const.}$ with $\chi_0=3\beta/\alpha$ and 
$\sigma_0=\alpha^2/(18\beta)$. The last term in Eq.~(\ref{lsf}) was 
introduced to allow this solution. The above solution is stable, i.e., for a 
range of initial conditions one still finds the desired asymptotic behavior 
$\chi\sim t^{-2}$ and $\sigma\sim\mbox{const.}$ for $t\to\infty$. This is 
easy to check numerically setting, e.g., $\alpha=\beta=1$. The stability 
does not depend on the precise values of these parameters. 

Thus, an asymptotic decay of the energy densities associated with the 
fields $\chi$ and $\sigma$, which is sufficiently fast to be consistent with 
a FRW cosmology, is realized without any fine tuning. What remains to be 
done is the replacement of the various $t$-dependent functions by the 
dynamics of appropriate fields.

\section{Adding matter, quintessence, and gravitational dynamics}
To embed the above adjustment mechanism in a realistic universe that 
includes matter and gravitational dynamics, assume first that the energy 
densities associated with $\chi$ and $\sigma$ remain small compared to the 
total energy density throughout the evolution. The total energy density 
is taken to consist of $\sim 20\%$ dark matter and $\sim 80\%$ quintessence. 
The cosmon field can be used to realize the explicit $t$ dependence in 
Eq.~(\ref{lsf}). 

It is known that a system with matter and a scalar field $\phi$ that is 
governed by an exponential potential $V_Q(\phi)=e^{-a\phi}$ gives 
rise to a realistic late cosmology with a fixed ratio of matter and field 
energy densities~\cite{wet,rp,cds}. The differential equations describing 
the system read 
\bea
&&\ddot{\phi}+3H\dot{\phi}+V_Q'(\phi)=0\\
&&6H^2=\rho+\frac{1}{2}\dot{\phi}^2+V_Q(\phi)\\
&&\dot{\rho}+3H\rho=0\,,
\eea
where $\rho$ is the density of dark matter. One finds the stable solution 
$H=(2/3)\,t^{-1}$, $\phi=(2/a)\ln t$ and $\rho=\rho_0 t^{-2}$. For $a^2=2$ 
one has $\rho_0=2/3$, which corresponds to a realistic dark matter to 
quintessence ratio. The explicit time dependence in Eq.~(\ref{lsf}) can now 
be replaced by a coupling to $\phi$. Technically, this is realized by the 
substitution $t^2\to e^{a\phi}$ in ${\cal L}_{SF}$. 

Now the $\chi$--$\sigma$ system of the last section and the 
$\phi$--$\rho$--gravity system described above have to be combined and it 
has to be checked whether the stability of each separate system suffices 
to ensure the stability of the complete system.

The complete Lagrangian, including the curvature term and the effective 
standard model action, Eq.~(\ref{ssmf}), reads
\be
{\cal L}=R+\frac{1}{2}(\partial\chi)^2F(\sigma,\phi)+\frac{1}{2}
(\partial\sigma)^2+\frac{1}{2}(\partial\phi)^2+V(\chi,\sigma,\phi) \,,
\ee
where
\be F(\sigma,\phi)=\sigma^2\,e^{2a\phi}\qquad\mbox{and}\qquad V(\chi,\sigma, 
\phi)=\alpha\sigma\chi+(1-\beta\sigma)\,e^{-a\phi}\,.
\ee
In a flat FRW universe, it gives rise to the equations of motion
\bea
&&\ddot{\chi}+(3H+\dot{F}/F)\dot{\chi}+\frac{1}{F}\,\frac{\partial V}
{\partial \chi}=0\\
&&\ddot{\sigma}+3H\dot{\sigma}+\frac{\partial}{\partial \sigma}\left(V-
\frac{1}{2}\dot{\chi}^2F\right)=0\\
&&\ddot{\phi}+3H\dot{\phi}+\frac{\partial}{\partial \phi}\left(V-
\frac{1}{2}\dot{\chi}^2F\right)=0\\
&&6H^2=\frac{1}{2}\dot{\chi}^2F+\frac{1}{2}\dot{\sigma}^2+\frac{1}{2}
\dot{\phi}^2+V+\rho\\
&&\dot{\rho}+3H\rho=0\,.
\eea
They have the asymptotic solution 
\be
\chi=\chi_0t^{-2}\,,\quad \sigma=\sigma_0\,,\quad \phi=\phi_0+(2/a)\ln t\,,
\quad \rho=\rho_0t^{-2}\,,\quad H=(2/3)t^{-1}\,.
\ee
For the parameters $\alpha=\beta=1$ and $a^2=2$, one finds
\be
\chi_0=\frac{3}{c}\,, \quad \sigma_0=\frac{1}{18c}\,, \quad \phi_0=
\frac{\ln c}{\sqrt{2}}\,, \quad \rho_0=\frac{5}{3}-\frac{1}{c}-\frac{1}
{6c^2}\,,
\ee
where $c=\left(1+\sqrt{11/9}\,\right)/2$. The clustering part of the energy 
density is $\rho/6H^2\simeq 0.21$. 

We have checked numerically that the above solution is stable, i.e.,
that a small variation of the initial conditions does not affect the 
asymptotic behavior for $t\to\infty$. The stability does not depend on 
the precise values of the parameters $\alpha$, $\beta$ and $a$. 

A small amount ($\sim 1\%$) of baryons can be introduced as a perturbation. 
In the present context, baryons are quite different from the above 
non-standard-model dark matter since, by virtue of Eq.~(\ref{fco}), they 
couple to the Brans-Dicke field $\sigma$. Thus, the baryonic energy density 
is $\sim\sigma\,n_B$, where $n_B$ is the number density of baryons. It has 
been checked that such an additional term does not affect the stability of 
the above solution. Solar system tests of the post-Newtonian approximation 
to general relativity place an upper bound on the coupling of baryons to 
almost massless scalar fields. In the present setting, the relevant coupling 
depends on $c$ whose present numerical value is not compatible with 
phenomenology in this respect. However, this can probably be avoided in a 
more carefully constructed model or by the ad-hoc introduction of a kinetic 
term for $\sigma$ that grows for large $t$. We also have not yet implemented 
the desirable slow decrease of $\rho/6H^2$, which would influence the 
detailed dynamics.

\section{Conclusions} 
In the present letter, a dynamical adjustment mechanism for the cosmological 
constant is constructed. It can be at work in a late FRW universe and 
ensures that the cosmological constant vanishes asymptotically. The 
existence of a working late-time adjustment mechanism is interesting 
because of its possible observational consequences. 

The field theoretic model that is used to realize this adjustment mechanism 
is generic but relatively complicated. In particular, it involves a 
Brans-Dicke field which, for the set of parameters used above, couples too 
strongly to standard model matter. However, there seems to be no reason 
why, with a different choice of parameters or scalar potentials, it should 
not be possible to avoid this phenomenological problem. A systematic 
investigation of such possibilities requires the analytic understanding 
of the stability of the system -- a task that appears to be relatively 
straightforward.

The beauty of adjustment mechanisms lies in the fact that they are 
independent of all the intricacies of the field theoretic standard model 
vacuum. As a new ingredient~\cite{rub} we use time or, equivalently, the 
value of the cosmon field as an essential parameter. Recently suggested 
adjustment mechanisms with an extra dimension~\cite{led} are related to 
this idea since they also employ a new parameter, the position in the extra 
dimension, and adjust the cosmological constant to zero only at a certain 
value of this parameter. 

One may hope that this new approach to the construction of adjustment 
mechanisms will eventually lead to a completely realistic and testable 
model.

\end{document}